\documentclass[twocolumn,showpacs,english,prb,preprintnumbers,amsmath,amssymb,floatfix]{revtex4}
\usepackage{epsfig,psfrag}
\usepackage{dcolumn}
\usepackage{bm}
\usepackage{graphicx}
\usepackage{color}
\usepackage{esint}
\usepackage{babel}
\usepackage{xspace}
\newcommand{\vk}{\mathbf{k}}
\newcommand{\vvr}{\mathbf{r}}
\newcommand{\vq}{\mathbf{q}}
\newcommand{\vu}{\mathbf{u}}
\renewcommand{\Im}{\mathop{\mathrm{Im}}}

\begin{document}

\title{Electron-phonon scattering in topological insulator thin films}
\author{S{\'e}bastien Giraud}
\author{Arijit Kundu}
\author{Reinhold Egger}
\affiliation{Institut f\"ur Theoretische Physik, 
Heinrich-Heine-Universit\"at, D-40225  D\"usseldorf, Germany}
\date{\today}

\begin{abstract}
We present a theoretical study of electron-phonon scattering effects 
in thin films made of a strong topological insulator.  
Phonons are modelled by isotropic elastic continuum theory with 
stress-free boundary conditions, and the interaction with the 
helical surface Dirac fermions is 
mediated by the deformation potential.
We determine the temperature-dependent electrical resistivity $\rho(T)$ and the 
quasiparticle decay rate $\Gamma(T)$ observable in photoemission. 
The low- and high-temperature power laws for both quantities are
obtained analytically. Detailed estimates covering the full 
temperature range are provided for Bi$_2$Se$_3$. 
\end{abstract}

\maketitle

\section{Introduction}

The recently discovered state of matter called
``topological insulator'' (TI) currently represents 
one of the most active areas in condensed matter 
physics.\cite{hasan,qizang}  TIs are characterized
by an insulating gap in the bulk but at the same time have an odd
number of gapless surface modes protected against all time-reversal 
invariant (and sufficiently weak) perturbations.\cite{fukane,qiz}  
In a three-dimensional (3D) TI, these surface modes correspond to 
massless two-dimensional (2D) Dirac fermions, where 
the spin direction is in the surface plane and
perpendicular to momentum (``spin-momentum locking'').  A typical reference
material is Bi$_2$Se$_3$ with a bulk gap $\Delta_b\approx 0.3$~eV. 
The helical Dirac electron property of the TI surface state
has been experimentally confirmed by 
spin- and angle-resolved photoemission spectroscopy 
(ARPES).\cite{hasan,arpes}
Transport experiments are more difficult in that respect
 since the surface contribution
is often masked by the residual conductivity due to impurities or
defects in the bulk.\cite{butch,ong,analytis}  
In thin films made of TI materials,
however, the bulk contribution is largely suppressed relative to the 
surface contribution, rendering the latter easier to observe. 

In this paper, we provide a detailed theoretical analysis of 
both the temperature-dependent resistivity $\rho(T)$ and the
quasiparticle lifetime $\Gamma(T)$ 
(observable in ARPES\cite{eiguren,echenique}) for 
a thin TI film.  The approach taken here generalizes previous
work for the semi-infinite geometry (with only one surface) 
by two of us\cite{giraud} to the film geometry.  
This brings about several important changes
compared to Ref.~\onlinecite{giraud} that are discussed below.  
We model the electronic part by retaining
only the Dirac surface states obtained from the low-energy 
bandstructure,\cite{zhang1} and our theory always assumes  that the Fermi level
is located inside the bulk gap.
We note in passing that Ref.~\onlinecite{zhang1} provides 
more accurate parameter estimates as the earlier paper by the same 
authors,\cite{zhang0} and we here adopt their new parameters in
our calculations using Bi$_2$Se$_3$ as example.
A similar parameter set has been published in Ref.~\onlinecite{shan}.

In sufficiently thin films, the hybridization of the two surface states eventually causes insulating behavior, as has recently 
been observed experimentally from ARPES for Bi$_2$Se$_3$ films.\cite{zhang00,expins}
For Bi$_2$Se$_3$, several calculations predicted\cite{shan,liu,linder1} a gap $\Delta(L)$ with (as a function of the width $L$) oscillations superimposed on an exponential decay. Similar calculations, however, found no oscillations, with a well-established TI phase already for 
$L\ge 3$ quintuple layers (QLs).\cite{newdft,abinitio}
Using the parameter set of Ref.~\onlinecite{zhang1}, we also 
find no evidence for oscillations in $\Delta(L)$, see Sec.~\ref{sec2a} below. 
For large width, one then has (upper and lower) massless
Dirac fermion surfaces.\cite{franz} 

Our working assumption below is that electron-phonon scattering 
is the dominant source of quasiparticle decay and backscattering.
Electron-electron interactions are indeed expected to give only
subleading corrections to the resistivity as long as   
$T\agt 1$~mK.\cite{eeint} Disorder effects are more likely to compete with 
phonon-induced backscattering effects. However, for elevated temperatures,
$T\agt 100$~K, phonon effects dominate even for present-day samples,
and anticipating higher purity films in the future, this crossover 
temperature may be lowered significantly.  ARPES setups allowing to test 
our predictions for the quasiparticle decay rate are basically
available.\cite{arpes,park,hofmann2,pan}
Other surface scattering techniques have also been applied to
extract the phonon dispersions.\cite{chamon}
We here follow Ref.~\onlinecite{giraud} and model the phonons
using elastic continuum theory.\cite{landau}  Since even at room temperature,
one effectively probes low energy scales, we
keep only long-wavelength acoustic phonon modes. For these, 
previous work on related materials has shown\cite{jenkins,huang}
that isotropic elastic continuum theory provides a reasonable approximation.
The phonon eigenmodes in the thin film geometry and their
coupling to electronic modes
have previously been determined in the context of 
semiconductor quantum well structures.\cite{bannov} (Note that 
the semi-infinite case has been treated in Ref.~\onlinecite{sirenko}.)
We basically reproduce the phonon eigenmodes of Ref.~\onlinecite{bannov},
but the coupling to the helical electronic eigenstates in a TI film is
different from the semiconductor case.
Note that piezoelectric couplings are suppressed by symmetry 
here,\cite{huang} and spin-phonon type couplings\cite{thalmeier} are 
also expected to be subdominant 
to the deformation potential taken into account below.

Most TI experiments have so far addressed only optical phonons,\cite{shahil}
cf.~also the corresponding situation for Bi surfaces,\cite{hofmann}
but very recently ARPES studies reported phonon-induced broadening
of the lineshape in TIs.\cite{park,hofmann2,pan}  The 
observed Bi$_2$Se$_3$ electron-phonon coupling 
strength,\cite{hofmann2} which has been 
extracted from the prefactor in the high-temperature 
quasiparticle decay rate $\Gamma\propto T$, 
is in good agreement with our theoretical estimates.\cite{giraud}
This indicates that the low-energy 
approach indeed provides a reasonable starting  point.
To the best of our knowledge, no detailed measurements for 
the temperature dependence of the TI film resistivity have
 been reported  so far.  We mention in passing
that for the related case of a 2D graphene monolayer, a similar comparison
of theory\cite{dassarma,felix} to experiment\cite{kim} has turned
out to be successful. 
Remarkably, the electron-phonon coupling observed in Ref.~\onlinecite{hofmann2} and independently estimated by us\cite{giraud} turns out to be quite large.  Under room temperature conditions, the resulting lifetime of helical quasiparticles is therefore short, and the resistivity is rather large. 
This behavior is substantially different from what is found in graphene. We suspect that this is (partially) due to the different Debye temperatures in both materials.

The structure of the remainder of this paper is as follows.
In Sec.~\ref{sec2} we discuss the model for the surface states
in the thin film and their coupling to the quantized phonon modes.
We then turn to the calculation of the electrical resistivity
in Sec.~\ref{sec3}, followed by the lifetime broadening 
in Sec.~\ref{sec4}. The paper concludes with a brief discussion in
Sec.~\ref{sec5}.  Technical details of our calculations can be found in various 
appendices. Note that we use units with $\hbar=1$.

\section{Model}\label{sec2}

In this section we describe the model employed in our study of
electron-phonon scattering in a TI film.  The model parameters
below are chosen for Bi$_2$Se$_3$ as a concrete example.
The film has infinite extension in the $xy$ plane  and the
 width $L$, where $|z|<L/2$.  
We start by reviewing the construction of the effective
surface Hamiltonian describing the (upper and lower) electronic 
surface states of a TI film. 

\subsection{Electronic surface states}
\label{sec2a}

Keeping all terms up to second order in the momentum
around the $\Gamma$ point, $(k_x,k_y,k_z)$, 
the low-energy physics of 3D TI materials  
like Bi$_2$Se$_3$ or Bi$_2$Te$_3$ is well described by 
an effective four-band model.\cite{hasan}  Using the basis states
$\{ \mbox{$|P1^+_z,\uparrow\rangle$}, |P2^-_z,\uparrow\rangle, 
|P1^+_z,\downarrow\rangle,|P2^-_z,\downarrow\rangle \}$, 
the low-energy bulk Hamiltonian reads\cite{zhang1,zhang0,shan}
\begin{equation}\label{model}
H=\begin{pmatrix} \epsilon_0+ M & -i A_{1}k_{z} & 0 & A_{2}k_{-} \\
iA_{1}k_{z} &\epsilon_0 -M & A_{2}k_{-} & 0 \\
0 & A_{2}k_{+} & \epsilon_0+M & -iA_{1}k_{z} \\
A_{2}k_{+} & 0 & i A_{1}k_{z} & \epsilon_0-M 
   \end{pmatrix}
\end{equation}
with $\epsilon_{0}=C+D_{1}k_{z}^2+D_{2} (k_x^2+k_y^2)$,
$M =M_{0}-B_{1}k_{z}^2-B_{2}(k_x^2+k_y^2)$ and $k_\pm=k_x\pm i k_y$. 
The model parameters for Bi$_2$Se$_3$ have been determined from 
first principles,\cite{zhang1} 
\begin{eqnarray}\label{parameters}
M_0 &=& 0.28~\mbox{eV}, \quad C=-0.0083~ \mbox{eV},\\ \nonumber
A_1 &=& 2.26~\mbox{eV\AA}, \quad A_2 = 3.33~\mbox{eV\AA},  \\ \nonumber
B_1 &=& 6.86~\mbox{eV\AA}^2 , \quad B_2=44.5~\mbox{eV\AA}^2,\\ \nonumber
D_1 &=& 5.74~\mbox{eV\AA}, \quad D_2 = 30.4~\mbox{eV\AA}^2.
\end{eqnarray}
We may write the Hamiltonian \eqref{model} in the form $H=H_0+H'$, where
$H_0 = \begin{pmatrix} h_0(k_z) & 0 \\ 0 & h_0(k_z) \end{pmatrix}$
is the $2\times 2$ block matrix obtained for $k_x=k_y=0$, with
\begin{equation}
h_0(k_z)=\begin{pmatrix}  \epsilon_0(k_z) + M_0-B_1 k_z^2 & -iA_1k_z\\
iA_1k_z & \epsilon_0(k_z)-M_0+B_1 k_z^2 \end{pmatrix}.
\end{equation}
Note that eigenstates of $H_0$ have conserved spin.

In order to find the surface states in the film geometry,
we follow the usual strategy\cite{zhang1,zhang0,shan}
and first look for general bispinor eigenstates of $h_0$,
\begin{equation}\label{h0eq}
h_0(k_z\to -i\partial_z) \Psi(z) = E_0 \Psi(z).
\end{equation}
The general solution to Eq.~\eqref{h0eq} reads ($j=\pm,s=\pm$)
\begin{equation}\label{psidef}
 \Psi(z) =\sum_{js} c_{js} e^{-s\eta_j z} \begin{pmatrix}
    E_0-C+M_0+(D_1+B_1)\eta_{j}^2\\ -sA_1\eta_{j}
   \end{pmatrix}
\end{equation}
with arbitrary $c_{js}$ and the inverse lengthscales 
\[
\eta_\pm  = \left[ \left(-\tilde{B}\pm
\sqrt{\tilde{B}^2-4\tilde{A}\tilde{C}}\right)/(2\tilde{A})\right]^{1/2},
\]
where $\tilde{A}= D_1^2-B_1^2, 
\tilde{B} = A_1^2-2[M_0B_1+D_1(C-E_0)],$ and
$\tilde{C} = (E_{0}-C)^2-M_{0}^2$.
The Dirichlet boundary conditions 
defining the film geometry, $\Psi(z=\pm L/2)=0$, then
imply the transcendental equation
\[
\frac{[E_{0}-C+M_0+(D_1+B_1)\eta_+^2]\eta_-}
{[E_{0}-C+M_0+(D_1+B_1)\eta_-^2]\eta_+} =
 \frac{\tanh(\eta_-L/2)}{\tanh(\eta_+ L/2)}, 
\]
or the same condition with $\eta_+\leftrightarrow\eta_-$ on the 
right hand side. Numerical solution of these equations yields
the $\Gamma$ point energies $E_0^{(\pm)}$. The corresponding 
eigenstates $\Psi_\pm(z)$ follow from Eq.~\eqref{psidef},
\begin{equation}\label{surface}
\Psi_\pm(z) = {\cal N}_\pm \begin{pmatrix} (D_1+B_1) \Lambda_\pm
F^\pm_\mp(z) \\ A_1 F^\pm_\pm (z) \end{pmatrix},
\end{equation}
where the ${\cal N}_\pm$ are normalization constants and  
\[
\Lambda_\pm= \left[ \frac{\eta_+^2-\eta_-^2}{\eta_+\coth^\pm(\eta_+L/2)-
\eta_-\coth^\pm(\eta_-L/2)} \right]_{E_0^{(\pm)}}
\]
with $\coth^+(y)=\coth(y)$ and $\coth^-(y)=\tanh(y)$.
Finally, the $F$ functions are
\[
F^\pm_+(z) = \left [ \frac{\cosh(\eta_+ z)}{\cosh(\eta_+L/2)}
-\frac{\cosh(\eta_-z)}{\cosh(\eta_-L/2)} \right]_{E_0^{(\pm)}},
\]
where $F^\pm_-$ follows with $\cosh\to \sinh$.
Note that the eigenstates $\Psi_{\pm}(z)$ describe both
spin directions ($\sigma=\uparrow$ and $\sigma=\downarrow$).

We now project the full Hamiltonian $H$ to the basis spanned by
the surface states (\ref{surface}).  
We define Pauli matrices $\tau_{\alpha=x,y,z}$
switching between the two solutions $\Psi_{\tau=\pm}(z)$,
Pauli matrices $\sigma_\alpha$ in spin space, and use
$\tau_0$ and $\sigma_0$ as identities.  With the energy scales
\begin{equation}\label{e0del}
E_0=\frac{E_0^{(+)}+E_0^{(-)}}{2},\quad \Delta= E_0^{(+)}-E_0^{(-)},
\end{equation}
the low-energy (``surface'') 
Hamiltonian resulting from this projection reads
\begin{equation}\label{heff}
H_{\rm eff}= E_0\tau_0\sigma_0 +\frac{\Delta}{2}\tau_z\sigma_0
-A_2 W \tau_x (k_x\sigma_x+k_y\sigma_y) + {\cal O}(\vk^2),
\end{equation}
where $W=\langle \Psi_+|\Psi_-\rangle$.  
The parameter $\Delta(L)$ is precisely the surface state gap described in the 
Introduction.  For the parameters \eqref{parameters}, 
$\eta_-$ is always real.  However, $\eta_+$ is real for large $L$ but
purely imaginary for small $L$. In any case, we find
that $W$ is always real and positive.

Noting that $H_{\rm eff}$ commutes with $\tau_z\sigma_z$,
it can readily be diagonalized
by the unitary transformation $U({\vk})={\rm diag}(U_+,U_-)$, where
$\vk=(k_x,k_y)$ and the $U_{\upsilon=\pm}(\vk)$ are $2\times 2$ matrices 
in spin space, with $\upsilon$ denoting the eigenvalue of $\tau_z\sigma_z$.
With $\tan\alpha=2A_2 W |\vk|/\Delta$ and $\tan\theta=k_y/k_x$, we find
\begin{eqnarray}\label{unitary}
U_{\upsilon=+} &=& \begin{pmatrix} e^{-i\theta/2}\cos(\alpha/2) & e^{-i\theta/2}
\sin(\alpha/2) \\ -e^{i\theta/2} \sin(\alpha/2) & e^{i\theta/2} 
\cos(\alpha/2) \end{pmatrix},\\
\nonumber U_{\upsilon=-} &=& \begin{pmatrix} -e^{-i\theta/2}\sin(\alpha/2) & 
e^{-i\theta/2} \cos(\alpha/2) \\ e^{i\theta/2} \cos(\alpha/2) & e^{i\theta/2} 
\sin(\alpha/2) \end{pmatrix}.
\end{eqnarray}
Switching to second-quantized notation, the eigenstates
of $H_{\rm eff}$ correspond to helical fermions with annihilation operator 
\begin{equation}\label{helical}
c_{\vk ,\upsilon s} = \sum_\sigma [U_\upsilon(\vk)]_{\sigma s}^* \
d_{\vk, \tau=\upsilon\sigma, \sigma},
\end{equation}
where $d_{\vk,\tau\sigma}$ annihilates a spin-$\sigma$ 
electron with in-plane momentum $\vk$ in the transversal
 state $\Psi_{\tau}(z)$.
The low-energy electronic Hamiltonian (including the chemical
potential $\mu$) then takes the final form
\begin{equation}\label{helectr}
H_{\rm el} = \sum_{\vk;\upsilon,s=\pm} 
\epsilon_{\vk, s}^{} c^\dagger_{\vk,\upsilon s}
c^{}_{\vk,\upsilon s},
\end{equation}
where the dispersion relation is
\begin{equation}\label{disp}
\epsilon_{\vk,\pm}=E_0-E_0^\infty-\mu \pm \frac{\Delta}{2}\sqrt{
1+ (2A_2W/\Delta)^2 \vk^2}.
\end{equation}
We here choose the zero of energy by setting $E_0^\infty=
C+D_1 M_0/B_1=\lim_{L\to \infty}E_0^{(\pm)}$.
For the parameters \eqref{parameters}, we find $E_0^\infty\simeq 0.22$~eV.
Moreover, for $L\to \infty$, 
the lengthscales $\eta_\pm^{-1}$ are given by
$\eta_+^{-1}\simeq 12.3$~\AA~and $\eta_-^{-1}\simeq 1.9$~\AA.
For $k L\gg 1$, the dispersion relation (\ref{disp}) is linear, with
Fermi velocity $v_F\simeq 2.77\times 10^5$~m$/$s.
Note that the index $s=\pm$ in Eq.~\eqref{helectr} does not correspond to spin anymore.

\begin{figure}
\centering
\includegraphics[width=\linewidth]{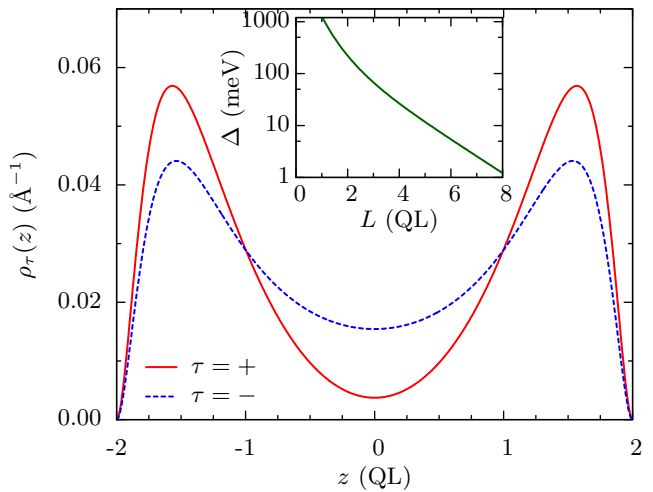}
\caption{\label{fig1} (Color online)
Electronic eigenstates for Bi$_2$Se$_3$ from 
Eqs.~\eqref{model} and \eqref{parameters}.
Main panel: Densities $\rho_{\tau}(z)$ in Eq.~\eqref{rdef}
for $L=4$~QL. Inset:  Gap $\Delta$ vs thickness $L$.
 Note the semi-logarithmic scale. }
\end{figure}

Similarly, 
the particle density operator $\hat n(\vvr,z)$ with $\vvr=(x,y)$ 
is written in terms of the $d_{\vk,\tau\sigma}$ operators,
\begin{equation}\label{dens}
\hat n (\vvr,z) = \sum_{\vk ,\vq,\tau, \sigma} e^{-i\vq \cdot \vvr} 
 \rho_{\tau}(z) d^\dagger_{\vk+\vq,\tau\sigma} d^{}_{\vk,\tau\sigma}.
\end{equation}
Using Eq.~\eqref{helical}, the density
operator (\ref{dens}) can be transformed to the helical basis.
We show the single-particle densities for the surface states [Eq.~\eqref{surface}],
\begin{equation}\label{rdef}
\rho_{\tau}(z)=\left[\Psi_{\tau}^\dagger \cdot \Psi_{\tau}\right](z),
\end{equation}
in Fig.~\ref{fig1} for a film thickness
of $L=4$~QL, where 1~QL~$\simeq 9.5$\AA~for Bi$_2$Se$_3$.\cite{apl}  
This demonstrates that already for quite thin films, 
Eq.~\eqref{surface} describes surface states.
Note that $\rho_\tau(z)$ is an even function of $z$.
The inset of Fig.~\ref{fig1} shows the numerically obtained gap $\Delta(L)$, 
demonstrating the absence of oscillatory behavior for the parameters 
\eqref{parameters} as well as the  exponential decay 
of $\Delta(L)$ due to the exponentially vanishing overlap of 
both surface states.   We note in passing that for the 
parameters in Ref.~\onlinecite{zhang0}, Eq.~\eqref{e0del} instead
predicts an oscillatory decay of $\Delta(L)$.

\subsection{Phonon model}
\label{sec2b}

We now discuss the long-wavelength acoustic phonon modes 
in the TI film.  We employ isotropic elastic continuum 
theory, where the longitudinal ($c_l$) and transverse ($c_t$) 
sound velocities correspond to the two Lam{\'e} constants.\cite{landau}
In Bi$_2$Se$_3$, they are given\cite{shoemaker,richter} by 
$c_l\simeq 2900$~m$/$s and $c_t\simeq 1700$~m$/$s, respectively.
Moreover, the mass density is $\rho_M=7680$~kg$/$m$^3$.\cite{wiese}
In order to model the film geometry, we impose
stress-free boundary conditions\cite{landau} at $z=\pm L/2$. 
The quantized phonon eigenmodes for this problem have been determined in 
Ref.~\onlinecite{bannov}.  For convenience, we briefly summarize 
the results next.  

Different phonon modes are labeled by a set of quantum numbers, 
$\Lambda=(\vq,\lambda,n)$, where $\vq=(q_x,q_y)$ is the surface momentum,
$\lambda\in (H,S,A)$ denotes the mode type, and $n\in \mathbb{N}$ 
is a branch index corresponding to the quantization of transverse momentum.
The horizontal shear mode ($\lambda=H$) decouples from all other modes
and does not generate a deformation potential,\cite{bannov} and we 
do not discuss this mode further.  We are left with
transversally symmetric (dilatational, $\lambda=S$) and antisymmetric  
(flexural, $\lambda=A$) phonons.
Denoting the dispersion relation of a given phonon mode $\Lambda$ by
$\Omega_\Lambda$ (see below) and the surface area  by ${\cal A}$,  
the displacement field operator is 
\begin{equation}\label{udef}
\mathbf{U}(\vvr,z,t) = \sum_\Lambda 
\frac{e^{i(\vq \cdot \vvr-\Omega_\Lambda 
t)}}{\sqrt{2\rho_M {\cal A} \Omega_\Lambda}} \ \vu_\Lambda(z) 
\ b_\Lambda + \mathrm{H.c.},
\end{equation}
where $b_\Lambda$ is a bosonic annihilation operator and
the noninteracting phonon Hamiltonian is
\begin{equation}\label{freephon}
H_{\rm ph}=\sum_\Lambda \Omega_\Lambda \left(b_\Lambda^\dagger b_\Lambda^{}
+ 1/2 \right).
\end{equation}
The orthonormal eigenmodes $\vu_\Lambda(z)$ in Eq.~\eqref{udef} describe linear 
combinations of $e^{\pm i k_{l,t} z}$ waves, where
\begin{equation}\label{eq:klkt}
k_{l,t} = \sqrt{(\Omega_\Lambda/c_{l,t})^2-q^2};
\end{equation}
$k_{l,t}=i\kappa_{l,t}$ with 
$\kappa_{l,t}=\sqrt{q^2-(\Omega_\Lambda/c_{l,t})^2}$ for 
$\Omega_\Lambda<c_{l,t}q$. Writing $\vu_\Lambda(z)$ in the form
\begin{equation}\label{nuz}
\vu(z) = \left(iq\phi_l-\frac{d\phi_t}{dz} \right) \hat e_q + 
\left( \frac{d\phi_l}{dz} +iq \phi_t \right)\hat e_z,
\end{equation}
where $\hat e_q=\vq/q$ and
\begin{equation}\label{philt}
\phi_{l,t} = a_{l,t}\cos(k_{l,t}z) + b_{l,t} \sin(k_{l,t}z),
\end{equation}
the stress-free boundary conditions at $z=\pm L/2$ yield
\begin{eqnarray}\label{bcs}
2iq\frac{d\phi_l}{dz}-(q^2-k_t^2)\phi_t & = & 0, \\ \nonumber
2iq\frac{d\phi_t}{dz}+(q^2-k_t^2)\phi_l & = & 0.
\end{eqnarray}
Since both equations have to be fulfilled at $z=\pm L/2$, we have
four linear equations for the four 
unknown parameters $(a_{l,t},b_{l,t})$.  Setting the corresponding
determinant to zero, we obtain the following two possibilities.
First, for  symmetric modes ($\lambda=S$), we have the condition
\begin{eqnarray}\label{condS}
 (q^2-k_t^2)^2\cos(k_lL/2)\sin(k_tL/2) &+& 
\\ \nonumber 4q^2k_lk_t\sin(k_lL/2)\cos(k_tL/2)& = & 0.
\end{eqnarray}
Numerical solution of this transcendental equation
obtains the quantized set of dilatational phonon 
frequencies $\Omega_{\Lambda=(\vq,S,n)}$.
The corresponding eigenvector, $\vu_\Lambda(z)$, 
follows from Eqs.~(\ref{nuz}) and \eqref{philt} with $a_t=b_l=0$ and 
\begin{equation}\label{abS}
a_l=\frac{2{\cal N}_S q}{\cos(k_l L/2)},\quad
b_t= \frac{i{\cal N}_S (q^2-k_t^2)}{k_t \cos(k_t L/2)}.
\end{equation}
Second, for antisymmetric modes ($\lambda=A$), we arrive again
at the condition in Eq.~\eqref{condS} but with the exchange 
$\cos\leftrightarrow \sin$.  Solving that equation yields the
set $\Omega_{\Lambda=(\vq,A,n)}$ of quantized flexural phonon modes.
The eigenvector $\vu_\Lambda(z)$ follows again from 
Eqs.~\eqref{nuz} and \eqref{philt}, where now $a_l=b_t=0$ and
\begin{equation}\label{abA}
b_l=\frac{2{\cal N}_A q}{\sin(k_l L/2)},\quad
a_t= \frac{-i {\cal N}_A (q^2-k_t^2)}{k_t \sin(k_t L/2)}.
\end{equation}
The normalization factors ${\cal N}_{\lambda=S,A}$ appearing in 
Eqs.~\eqref{abS} and \eqref{abA} are given in Appendix \ref{appa}. 

\begin{figure}
\centering
\includegraphics[width=\linewidth]{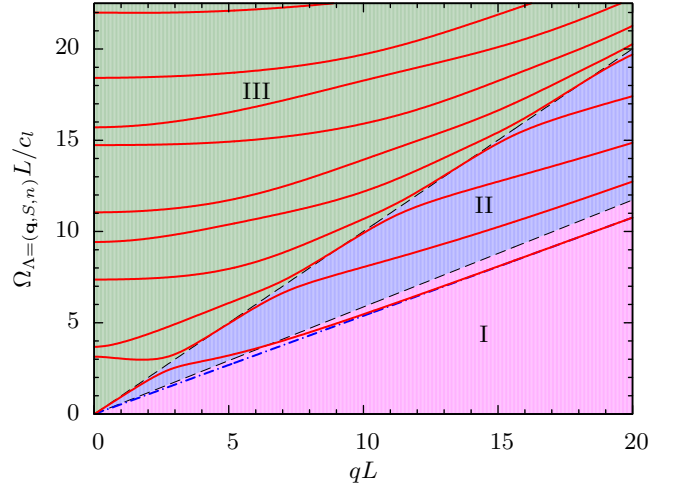}
\caption{\label{fig2} (Color online)
Phonon dispersion relation, $\Omega_\Lambda$ vs $q$, 
for the symmetric ($\lambda=S$) mode (red solid curves).
Shown are the ten lowest branches corresponding to the index $n$.
Dashed lines separate regions I, II, and III (see main text).
The dash-dotted line gives the dispersion relation in Eq.~(\ref{rayleigh}); 
note that the $n=1$ mode coincides with the Rayleigh mode for $qL\gg 1$.  }
\end{figure}

Numerical solution of Eq.~\eqref{condS} yields 
the spectrum, $\Omega_\Lambda$, for the symmetric mode ($\lambda=S$).
The result is shown in Fig.~\ref{fig2}.  We
 distinguish three different regions, namely a case
where both $k_l$ and $k_t$ are purely imaginary (region I), 
when only $k_l$ is purely imaginary but $k_t$ is real (region II), and 
finally a case where both $k_l$ and $k_t$ are real (region III).
We observe from Fig.~\ref{fig2} that the $n=1$
mode is the finite-width analogue
of the well-known Rayleigh surface mode.\cite{landau,sirenko} 
For the semi-infinite geometry, the Rayleigh mode is
the lowest-lying phonon.\cite{footcorr} It has the dispersion relation 
\begin{equation} \label{rayleigh}
\Omega=c_R q,\quad c_R\simeq 0.92 c_t.
\end{equation}
In fact, for $qL\gg 1$, both Eq.~\eqref{condS} and the 
corresponding equation for $\lambda=A$ reduce to 
\[
(q^2+\kappa_t^2)^2=4q^2\kappa_t\kappa_l.
\]
As discussed in Ref.~\onlinecite{landau}, this equation 
readily yields the sound velocity $c_R$ of the Rayleigh mode.

\subsection{Electron-phonon coupling}\label{sec2c}

The dominant coupling of the above phonon modes to
the electronic surface states comes from
the deformation potential,\cite{giraud} which couples the local electronic
density $\hat n(\vvr,z)$  [Eq.~\eqref{dens}] to the divergence
of the displacement vector, $\nabla\cdot \mathbf{U}(\vvr,z)$, see
Eq.~\eqref{udef}.  Since the surface state density 
$\rho_\tau(z)$ in Eq.~\eqref{rdef} is even in $z$, 
the antisymmetric phonon mode ($\lambda=A$) does not
couple to the surface states. 
We therefore keep only the symmetric phonon mode 
from now on (and omit the index $\lambda=S$). 
Transforming Eq.~\eqref{dens} to the helical basis, see Eq.~\eqref{helical},
the second-quantized electron-phonon coupling Hamiltonian reads
\begin{equation}\label{heph}
H_{{\rm e-ph}}= \frac{\alpha}{\sqrt{{\cal A}}} \sum_{\vq,\vk,n;\upsilon,s,s'} 
M_{\vk,\vq,n}^{(\upsilon,s,s')} b_{\vq,n} 
c^\dagger_{\vk+ \vq,\upsilon s} c^{}_{\vk ,\upsilon s'} + {\rm H.c.},
\end{equation}
where the $M$ matrix elements involve the unitary 
matrices $[U_\upsilon(\vk)]_{s\sigma}$ in Eq.~\eqref{unitary},
\begin{widetext}
\begin{equation}\label{ephcoupl}
M_{\vk,\vq,n}^{(\upsilon,s,s')}= - \frac{1}{\sqrt{2\rho_M\Omega_{q,n}}} 
\left( \frac{\Omega_{q,n}}{c_l}\right)^2 
\sum_{\sigma} [ U_{\upsilon} (\vk+\vq)]^*_{s\sigma} 
[ U_{\upsilon}(\vk) ]_{\sigma s'}
\int_{-L/2}^{L/2}dz\ \rho_{\tau=\upsilon\sigma}(z) \phi_l(z),
\end{equation}
\end{widetext}
with the phonon dispersion $\Omega_{q,n}$ in Fig.~\ref{fig2}; 
$\phi_l$ is given by Eqs.~\eqref{philt} and \eqref{abS}. 
The deformation potential strength $\alpha$ in Eq.~\eqref{heph}
can be estimated as follows.  The high-temperature behavior of the
on-shell imaginary part of the electronic self-energy is (see Sec.~\ref{sec4})
\begin{equation}\label{imsigma}
\Im\Sigma(k,T)= -\pi \lambda_k k_B T,
\end{equation}
which allows to experimentally extract the dimensionless
effective electron-phonon coupling constant 
$\lambda_k$.  The relation (\ref{imsigma}) has been observed 
for Bi$_2$Se$_3$ in ARPES experiments,\cite{hofmann2} and $\lambda=
0.25\pm 0.05$ has been measured. In these experiments, 
the Fermi level was near the bottom of the conduction band,  
$\mu\simeq 0.28$~eV, and $k$ in Eq.~\eqref{imsigma}
corresponds to energies $\approx 50$ to 100 meV above the Dirac point. 
Computing $\lambda_k$ within our model, see Sec.~\ref{sec4}, 
the observed value for $\lambda$ corresponds to 
$\alpha=(30\pm 8)$~eV.  We employ the value $\alpha=30$~eV below.  

The total Hamiltonian employed in the following sections is then
given by $H=H_{\rm el}+H_{\rm ph}+H_{\rm e-ph}$, see Eqs.~\eqref{helectr},
\eqref{freephon} and \eqref{heph}.  We first address the phonon-induced
 resistivity $\rho$ in Sec.~\ref{sec3} and then turn to the quasiparticle 
lifetime in Sec.~\ref{sec4}.

\section{Resistivity}\label{sec3}

Here we discuss the $T$-dependent phonon contribution
to the electrical resistivity, $\rho$, in the TI film, using the 
Hamiltonian described in Sec.~\ref{sec2}. 
As explained in Sec.~\ref{sec2c}, only symmetric (dilatational)
phonon modes can cause a finite resistivity for the low-energy surface
states within the bulk gap.  
We compute $\rho$ within the framework of the linearized
Boltzmann equation,\cite{ashcroft} 
which has also been employed previously for the related
graphene case.\cite{dassarma,felix}
The resulting quasiclassical estimate for $\rho$ is 
valid\cite{dassarma} as long as $\rho$ is small 
compared to the resistance quantum,
 $\rho\ll h/e^2\simeq 25.8$~k$\Omega$.
We sketch the standard derivation\cite{bannov,dassarma,felix,dassarma2}
for $\rho$ in Appendix \ref{appb}. The result takes the form
\begin{equation}\label{eq:rho}
\frac{1}{\rho}=\frac{e^2}{2} \sum_{\upsilon,s=\pm}
\int \frac{d\vk}{(2\pi)^2}\,
v_{\vk, s}^2 \tau_\upsilon(\epsilon_{\vk,s}) \left[ -\partial_\epsilon 
n_F(\epsilon_{\vk, s}) \right],
\end{equation}
where the dispersion relation for helical fermions [Eq.~\eqref{disp}]
defines the group velocity, 
$v_{\vk, s}= \hat e_k\cdot \nabla_{\vk} \epsilon_{\vk,s}$.
Moreover, $n_F(\epsilon)$ is the Fermi function, and the 
energy-dependent electron-phonon transport scattering rate (inverse time) is
\begin{widetext}
\begin{equation}\label{momentum}
\frac{1}{\tau_\upsilon(\epsilon_{\vk,s})}=
\sum_{\vq,s'} \left(1-\frac{v_{\vk+\vq,s'}}{v_{\vk,s}}
\cos\theta_{\vk,\vq}\right) 
\frac{1-n_F(\epsilon_{\vk+\vq,s'})}{1-n_F(\epsilon_{\vk,s})}
W_{(\vk,\upsilon s)\to(\vk+\vq, \upsilon s')} ,
\end{equation}
where $\theta_{\vk,\vq}$
is the angle  between $\vk$ and $\vk+\vq$, and the transition
probabilities are obtained from Fermi's golden rule.
Using Eq.~\eqref{heph}, we find
\begin{equation}\label{transition}
W_{(\vk,\upsilon s)\to(\vk+\vq,\upsilon s')}=
\frac{2\pi \alpha^2}{\mathcal{A}} 
\sum_{n;\nu=\pm}  \nu n_B \left(\nu \Omega_{q,n}\right ) \
\left| M_{\vk,\vq,n}^{(s',s)}\right|^2 \ 
\delta\left(\epsilon_{\vk,s}+\nu \Omega_{q,n}-
 \epsilon_{\vk+\vq,s'} \right)  ,
\end{equation}
\end{widetext}
where $n_B(\epsilon)$ is the Bose function.
While the $M$ matrix elements (\ref{ephcoupl}) 
depend on the index $\upsilon=\pm$, we note that $|M|^2$ and therefore
the transition probabilities $W$ are $\upsilon$-independent.
This also implies that $\tau_\upsilon$ does actually not depend on $\upsilon$.

With the polar angle $\theta$ between $\vk$ and $\vq$, such that
\begin{equation}\label{thetadef}
\cos\theta_{\vk,\vq}= \frac{ k+q\cos\theta}
{\sqrt{k^2+q^2+2kq\cos\theta}},
\end{equation}
the angular integration in Eq.~\eqref{momentum} can be encapsulated in  
the ``transport Eliashberg function'', 
see also Ref.~\onlinecite{giraud}, 
\begin{widetext}
\begin{equation}\label{eq:F}
 {\cal F}_{k,n,s}^{(\nu)}(q)=\sum_{s'}\int_{-\pi}^\pi
\frac{d\theta}{2\pi} \left[1-\frac{v_{\vk+\vq,s'}}
{v_{\vk, s}}\cos\theta_{\vk,\vq}\right]
\left|M_{\vk,\vq,n}^{(s',s)}\right|^2 
\delta\left(\epsilon_{\vk ,s}+\nu \Omega_{q,n}- 
\epsilon_{\vk+\vq,s'}\right).
\end{equation}
This allows us to write the momentum relaxation rate (\ref{momentum})
in the form
\begin{equation} \label{eq:tau-F}
\frac{1}{\tau(\epsilon_{\vk ,s})}= 
\alpha^2\sum_{n,\nu}\int_0^{\infty} q dq \,
{\cal F}_{k,n,s}^{(\nu)}(q)\,\nu n_B(\nu \Omega_{q,n})
\frac{1-n_F(\epsilon_{\vk s}+\nu \Omega_{q,n})}
{1-n_F(\epsilon_{\vk s})}.
\end{equation}
\end{widetext}
The $\theta$-integration in Eq.~\eqref{eq:F} can then be carried out 
analytically. We quote the (lengthy) result in Appendix \ref{appc},
which is useful when computing ${\cal F}$ numerically.
For low temperatures, the quasi-elastic approximation, 
$ \Omega_{q,n}\ll \sqrt{(\Delta/2)^2+(A_2 W k)^2},$
is applicable and allows to simplify the full result for 
${\cal F}$ to the $\nu$-independent form 
\begin{eqnarray}\label{quasielastic}
{\cal F}_{k,n,s}(q) &=&  \Theta\left(2k-q\right) 
\frac{1}{\pi \sqrt{(2k/q)^2-1}}  \\ \nonumber &\times&
\frac{\sqrt{(\Delta/2)^2+(A_2 W k)^2}}{(A_2 W k)^2}
\left.\left| M_{\vk,\vq,n}^{(s,s)}\right|^2
\right|_{\theta_0} ,
\end{eqnarray}
where $\theta=\theta_0$ (see Appendix \ref{appc}) 
determines the polar angle between $\vk$ and $\vq$ 
appearing in the matrix element $M$,
and the Heaviside function is denoted by $\Theta(y)$.
Note that there is no contribution from interband transitions at low
temperatures.

The crossover temperature from the low- to the high-temperature
behavior in this system is set\cite{giraud,dassarma} 
by the Bloch-Gr\"uneisen temperature,
\begin{equation}\label{tbg}
T_{\rm BG}= 2 k_F c_R/k_B,
\end{equation}
with the Rayleigh velocity $c_R$ in Eq.~\eqref{rayleigh}.  
$k_F(L)$ is defined by $\epsilon_{k_F,s=+}=0$ with the dispersion
relation (\ref{disp}).
For $T\ll T_{\rm BG}$, the ${\cal F}$ function can be 
approximated by the quasi-elastic 
expression [Eq.~\eqref{quasielastic}].  It receives the dominant
contribution from the $n=1$ branch corresponding
to the Rayleigh surface phonon.  For small $q$,
we find $\Omega_{q,n=1}=c_s q$ with $c_s= 2754$~m$/$s (which is 
slightly below $c_l$), see also
Fig.~\ref{fig2}.  In addition, we have $\phi_l(z) = 2(c_t/c_s)^2/(q\sqrt{L})$ 
and $c_s\ll {\rm min}(|v_{k_F}|,A_2 W)$, leading to 
\[
{\cal F}_{k_F,1,\pm}(q)= \frac{(c_t/c_l)^4 }{\pi \rho_M
|v_{k_F}| c_s k_F^2 } \frac{q^2}{L}.
\]
This allows us to perform all remaining integrations and yields a $T^4$ law
for the resistivity at low temperatures,
\begin{equation}\label{rho:lowT}
\rho(T\ll T_{\rm BG}) =  \frac{h}{e^2} A \left(\frac{T}{T_{\rm BG}}\right)^4,  
\end{equation}
where the dimensionless prefactor $A$ is 
\begin{eqnarray}\label{Adef}
A &=& \frac{8 \gamma k_F \alpha^2}{\pi\rho_M v_{k_F}^2 c_s}  
\left(\frac{c_t c_R}{c_l c_s}\right)^4 \frac{1}{L},\\ \nonumber
\gamma &=& \left[\int_{-\infty}^\infty dx \frac{2e^x}{[(\pi^2+x^2) (e^x+1)]^2}
\right]^{-1}\simeq  68.4295. 
\end{eqnarray}
For $L\to \infty$, $A$ obviously vanishes. This suggests that 
for elevated temperatures (but still $T < T_{\rm BG}$)
and finite $L$, the $T^4$ law is replaced by the $L$-independent
$\rho\propto T^5$ law found in Ref.~\onlinecite{giraud}.
We can estimate the crossover temperature $T_c$ as follows. 
For $T< T_{\rm BG}$, we expect an expansion of the form
\[
(e^2/h)\rho=A(T/T_{\rm BG})^4+ \frac{B}{4}(T/T_{\rm BG})^5,
\] 
with $A\propto 1/L$ in Eq.~\eqref{rho:lowT} and
the $L$-independent constant $B$ given in
Ref.~\onlinecite{giraud}.  
The crossover from the $T^4$ law (for $T \alt T_c$) to
the $T^5$ law (for $T_c \alt T < T_{\rm BG}$) thus happens around
the temperature $T_c=(4A/B) T_{\rm BG}$. 
This gives $T_c\simeq 0.14 T_{\rm BG}/(k_F L)$, which is
 independent of the chemical potential 
since $T_{\rm BG}\propto k_F$.  For $L=4$~QL, we obtain $T_c\approx 0.9$~K.
The $T^4$ law can thus only be observed for very thin and clean TI films.

In the opposite high-temperature limit, 
essentially all phonon branches indexed by $n$ contribute to 
the transport Eliashberg function \eqref{eq:F}, see Appendix C.
Then the relaxation rate $\tau^{-1}(\epsilon_{\vk,s})$ in Eq.~(\ref{eq:tau-F}) 
is basically a linear function of the energy.  Since the 
linear term does not contribute to $\rho$ 
after integration in Eq.~\eqref{eq:rho}, we obtain the approximation $1/\rho
\simeq (e^2/h) v_{k_F} k_F \tau(\epsilon=0)$, where Eq.~\eqref{eq:tau-F}
yields the linear high-temperature law
\begin{equation}\label{rho:highT}
\rho(T\gg T_{\rm BG}) =  \frac{h}{e^2} C \frac{T}{T_{\rm BG}}   
\end{equation}
with the dimensionless prefactor
\begin{equation}\label{defC}
C= \frac{2\alpha^2 c_R}{v_{k_F}} \sum_{n,\nu=\pm}\int_0^\infty
qdq \frac{{\cal F}_{k_F,n,+}^{(\nu)}(q)}{\Omega_{q,n}} .
\end{equation}

\begin{figure}
\centering
\includegraphics[width=\linewidth]{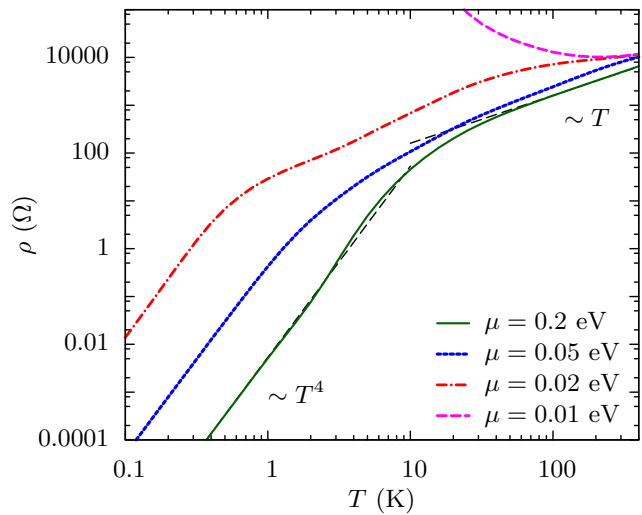}
\caption{\label{fig3} (Color online)
Phonon contribution to the resistivity $\rho$ vs temperature $T$
for a TI film of width $L=4$~QL and several values of the 
chemical potential $\mu$.  Dashed lines indicate the analytical
results for low [Eq.~\eqref{rho:lowT}] and high [Eq.~\eqref{rho:highT}]
temperatures. Note the double-logarithmic scale. }
\end{figure}

Next we show the full temperature dependence of $\rho$
obtained numerically for a fixed width $L=4$~QL and several
values of the chemical potential $\mu$, see Fig.~\ref{fig3}.
In that case, when measured relative
to $E_0^\infty$, we have $E_0^+\simeq 16$~meV and $\Delta/2 
\simeq 13$~meV.  For the lowest $\mu$ in Fig.~\ref{fig3},
the Fermi level is thus located inside the surface gap and 
one has a very large resistivity, where the quasi-classical
approach is not reliable in any case.
For low temperatures, $T<T_{\rm BG}$, the analytical result
\eqref{rho:lowT} with $\rho\propto T^4$ is nicely reproduced
by numerics.  In this temperature regime, only the Rayleigh
mode ($n=1$) is relevant, similar to what one
finds in the semi-infinite geometry.\cite{giraud} 
In the high-temperature limit, both the $\rho\propto T$
scaling and the prefactor $C$ in Eq.~\eqref{defC}
are also consistent with our numerical findings. 

\begin{figure}
\centering
\includegraphics[width=\linewidth]{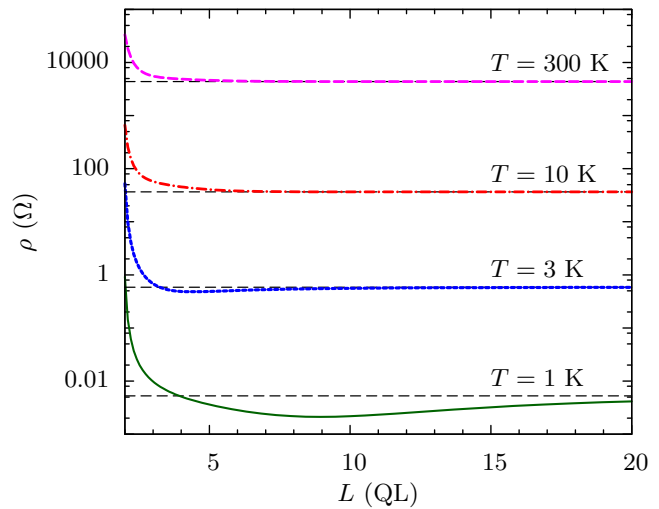}
\caption{\label{fig4} (Color online)
Width ($L$) dependence of the phonon contribution to the resistivity $\rho$
for $\mu=0.2$~eV and several temperatures. The dashed horizontal line
indicates one-quarter of the resistivity $\rho_\infty(T)$  in
the semi-infinite geometry with otherwise identical parameters.\cite{giraud}
}
\end{figure}

Finally, Fig.~\ref{fig4} shows the width ($L$) dependence of $\rho$ at fixed
chemical potential and for several $T$.  Two noteworthy observations 
can be drawn from Fig.~\ref{fig4}:  
First, for low temperatures we observe a ``dip'' in Fig.~\ref{fig4},
where $\rho(L)<\rho(L\to \infty)$ for intermediate values of $L$.
Second, for $L\to \infty$, $\rho(L)$ approaches $1/4$ of the
single-surface value $\rho_\infty(T)$ obtained for the semi-infinite
geometry.\cite{giraud}  Naively, we would expect $\rho(L\to\infty)
=\rho_\infty/2$ because of the presence of two surfaces in the film 
geometry.  This discrepancy indicates that the $L\to\infty$ limit
is singular, and it is not
possible to really decouple both surfaces in such an 
interacting system; see also Ref.~\onlinecite{franz}  for a related
discussion.

\section{Lifetime broadening} \label{sec4}

\begin{figure}
\includegraphics[width=\linewidth,angle=0]{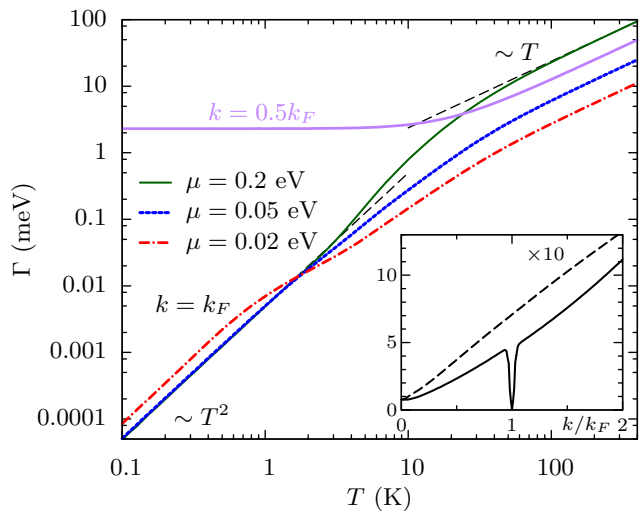}
\caption{\label{fig5} (Color online) Main panel: $T$-dependence of the
 decay rate $\Gamma$ of a TI film of width $L=4$~QL for $k=k_F$ and 
$k=0.5 k_F$.  For $k=0.5 k_F$, only the
$\mu=0.2$~eV result is displayed.  Dashed lines indicate the 
low- and high-temperature laws ($\Gamma\propto T^2$ and $\propto T$), 
respectively. Inset: $k$-dependence
of $\Gamma$ for $\mu=0.2$~eV and two different temperatures: $T=3$~K (solid line) and $T=300$~K (dashed line;
the shown result has to be multiplied by 10).  
 }
\end{figure}

\begin{figure}
\centering
\includegraphics[width=\linewidth]{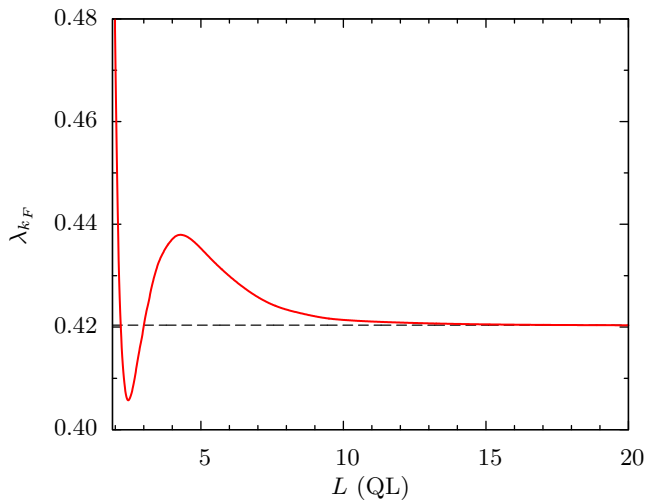}
\caption{\label{fig6} (Color online)
Width ($L$) dependence of the effective electron-phonon coupling constant at the Fermi level $\lambda_{k_F}$ for $\mu=0.2$~eV. The dashed horizontal line
indicates one-half of the effective coupling constant in
the semi-infinite geometry with otherwise identical parameters.\cite{giraud}
}
\end{figure}

Next we discuss the quasiparticle lifetime (inverse decay rate) 
for the surface fermions in the TI film due to their coupling to phonons,
see $H_{\rm e-ph}$ in Eq.~\eqref{heph}, which 
implies a finite  linewidth of ARPES 
spectral features.  The decay rate, $\Gamma_{k}(T) =-2 \Im\Sigma$, 
follows from the imaginary part
of the on-shell self-energy $\Sigma_{s=+}
(\vk,\omega=\epsilon_{\vk,s=+})$.\cite{foot}
Expanding up to second order in $H_{\rm e-ph}$, the ``rainbow'' diagram 
yields the self-energy
\begin{eqnarray}\label{selfe}
\Sigma_{s}(\vk,\omega) &=& \alpha^2 \sum_{n,s'}\int \frac{d\vq}{(2\pi)^{2}}
\left| M_{\vk,\vq,n}^{(s',s)}\right|^2  \\ \nonumber &\times& 
\sum_{\nu=\pm}\nu \frac{n_B(\nu \Omega_{q,n})+n_F(\epsilon_{\vk+\vq,s'})}
{\omega+i0^++\nu\Omega_{q,n}-\epsilon_{\vk+\vq,s'}}.
\end{eqnarray}
Introducing the Eliashberg function $F_{k,n,s}^{(\nu)}(q)$ exactly as
the transport Eliashberg function ${\cal F}$ in
Eq.~\eqref{eq:F} but without the factor $[1- (v_{\vk+\vq,s'} /
v_{\vk,s} ) \cos\theta_{\vk,\vq}]$, the quasiparticle decay rate follows as
\begin{eqnarray}\label{gammadef}
\Gamma_{k}(T) &=& \alpha^2\sum_{n,\nu}\int_0^\infty q dq \ 
F_{k,n,+}^{(\nu)}(q) \\ \nonumber &\times& \left[
n_B(\Omega_{q,n})+n_F(\Omega_{q,n}+\nu\epsilon_{k,+}) \right].
\end{eqnarray} 
Expanding this result for high temperatures, $T\gg T_{\rm BG}$,
as in Sec.~\ref{sec3} yields, see also Eq.~\eqref{imsigma}, a linear
$T$-dependence,
\begin{eqnarray}\label{hight}
\Gamma_{k}(T\gg T_{\rm BG}) &=&2\pi\lambda_{k} k_B T, \\ \nonumber
\lambda_{k}&=&\frac{\alpha^2}{2\pi}\sum_{n,\nu}\int_0^\infty qdq\, 
\frac{F^{(\nu)}_{k, n,+}(q)}{\Omega_{q,n}}.
\end{eqnarray}
The $L$-dependence of $\lambda_k$ is shown for $k=k_F$ in Fig.~\ref{fig6}.
We observe an oscillatory dependence, with a saturation at one-half
of the corresponding semi-infinite result.

For low temperatures and $k=k_F$, the decay rate is dominated by
the $n=1$ phonon mode with $q\to 0$. After some algebra, we
find that this implies a $T^2$ law,
\begin{equation}
\Gamma_{k_F}(T\ll T_{\rm BG}) = \frac{4\pi(c_t/c_l)^4 (k_F c_R\alpha)^2}{
\rho_M|v_{k_F}| c_s^3 } \frac{1}{L} 
\left(\frac{T}{T_{\rm BG}}\right)^2.
\end{equation}
Again, when $T\agt T_c$, the $T^2$ law (which scales $\propto 1/L$)
competes with the $L$-independent $T^3$ law
found in Ref.~\onlinecite{giraud}, see Sec.~\ref{sec3}.  
Finally, when $k\ne k_F$ and $T\ll T_{\rm BG}$, 
the quasiparticle decay rate saturates at the finite value
\begin{equation}
\Gamma_{k\neq k_F}=\alpha^{2}\sum_{n}\int_0^\infty qdq\, 
\Theta(|\epsilon_{k+}|-\Omega_{q,n}) \ F^{(\nu)}_{k, n,+}(q).
\end{equation}
with $\nu={\rm sgn}(k_F-k)$.

Figure \ref{fig6} shows that the $L\to \infty$ limit of the decay
rate always tends to $\Gamma_\infty(T)/2$, where $\Gamma_\infty$ is the 
corresponding decay rate for the semi-infinite geometry.\cite{giraud}
This discrepancy with the naive expectation 
$\Gamma(L\to \infty)=\Gamma_\infty$ has the same origin as 
the anomalous factor $1/2$ appearing in the large-$L$ behavior
of the resistivity discussed in Sec.~\ref{sec3}.

\section{Conclusions} \label{sec5}

In this paper we have studied the effects of long-wavelength acoustic phonons
on the topologically protected surface fermions in topological insulator films.
Our model employs the established low-energy electronic Hamiltonian and
an isotropic elastic continuum approach for the phonons, with the
deformation coupling providing the dominant interaction mechanism. 
The electron-phonon coupling turns out to be surprisingly strong,
in accordance with recent ARPES results.\cite{hofmann2}

Using a quasiclassical approach, we have computed the temperature-dependent
resistivity of the film due to phonon backscattering, and found a 
linear $T$ dependence above the Bloch-Gr\"uneisen temperature.  In this
temperature regime, the phonon-induced resistivity can overcome the
disorder-induced ($T$-independent) contribution and should be observable
with present samples.  Similarly, the linear $T$ dependence of the 
quasiparticle decay rate found here is observable\cite{hofmann2} in 
ARPES experiments.  The low-temperature behaviors of the resistivity
and of the quasiparticle decay rate are probably  more difficult to 
observe.

An interesting extension of our work would be to include the effects
of a magnetic field.
Magnetotransport measurements in thin films were recently performed\cite{magn}
and found clear evidence for Landau level formation associated with the
massless Dirac fermions forming on both surfaces.  The observed 
broadening of the Landau levels was assigned to disorder and/or 
interaction effects, but at elevated temperatures, our analysis indicates
that electron-phonon interactions may be relevant as well.

\acknowledgments
We thank Philip Hofmann for useful discussions.
This work was supported by the Humboldt foundation and by
the SFB TR 12 of the DFG.

\appendix
\section{Normalization constants}
\label{appa}

Here we provide the normalization constants ${\cal N}_{S,A}$
appearing in Eqs.~\eqref{abS} and \eqref{abA} in Sec.~\ref{sec2b}. 
Specifically, we get these constants after some algebra from
\begin{widetext}
\begin{eqnarray*}
\mathcal{N}_S^{-2}&=& \frac{\Omega_\Lambda^2}{2c_t^2}
\left[ \frac{c_t^2}{c_l^2}\frac{4q^2 L}{\cos^2(k_lL/2)}
\left(1+\frac{\sin(k_lL)}{k_lL}\right)+\frac{(q^2-k_t^2)^2 L}
{k_t^2\cos^2(k_tL/2)}\left(1-\frac{\sin(k_tL)}{k_tL}\right) 
-8(q^2-k_t^2)\frac{\tan(k_tL/2)}{k_t}\right],\\
 \mathcal{N}_A^{-2}&=&\frac{\Omega_\Lambda^2}
{2c_t^2}\left[ \frac{c_t^2}{c_l^2}\frac{4q^2 L}{\sin^2(k_lL/2)}
\left(1-\frac{\sin(k_lL)}{k_lL}\right)+\frac{(q^2-k_t^2)^2 L}
{k_t^2\sin^2(k_tL/2)}\left(1+\frac{\sin(k_tL)}{k_tL}\right)
+8(q^2-k_t^2)\frac{\cot(k_tL/2)}{k_t}\right].
\end{eqnarray*}
\end{widetext}

\section{Linearized Boltzmann equation}
\label{appb}

The linearized Boltzmann equation has been derived for closely 
related problems before,\cite{bannov,dassarma,dassarma2,felix} and we here
follow those works and  briefly sketch the derivation of Eq.~\eqref{eq:rho}.
In the quasiclassical approximation, the quasiparticle distribution 
function $f(\vvr,\vk,t)$ (for simplicity, we here
 omit the $\upsilon$ and $s$ indices) obeys the well-known Boltzmann equation. 
In the absence of a force $\mathbf{F}=-e\mathbf{E}$ 
due to the external electric field, $f$ reduces to a Fermi function, $f=n_F(\epsilon_{\vk})$, and the collision integral vanishes. 
In the presence of the force, $f$ is expanded in terms of Legendre polynomials $P_n(\cos\alpha)$, where $\alpha$ is the angle between $\vk$ and $\mathbf{F}$. Keeping only terms
linear in $\mathbf{F}$, we have $f(\vk)=n_F(\epsilon_{\vk})+\cos(\alpha) f_1(\epsilon_{\vk})$. Using detailed balance, for given transition matrix elements $W_{\vk\to \vk^\prime}$, 
we obtain the linearized Boltzmann equation (LBE),
\begin{eqnarray*}
 & & \mathbf{F}\cdot \mathbf{v}_\vk\
\partial_\epsilon n_F(\epsilon_{\vk})=
\sum_{\vk^\prime}W_{\vk\to\vk^\prime}\\ 
& \times & \left[\frac{n_F(\epsilon_\vk)}
{n_F(\epsilon_{\vk^\prime})}\cos(\alpha^\prime) 
f_1(\epsilon_{\vk^\prime})-\frac{1-n_F(\epsilon_{\vk^\prime})}{1-n_F(\epsilon_\vk)}\cos(\alpha)
f_1(\epsilon_\vk)\right].
\end{eqnarray*}
Using the \textit{Ansatz} $f_1(\epsilon_\vk)=-\tau(\epsilon_\vk) v_\vk F \partial_\epsilon n_F(\epsilon_{\vk}),$
after some algebra  the LBE leads to the linear integral equation
\[
 \frac{1}{\tau(\epsilon_\vk)}=\sum_{\vk^\prime}W_{\vk\to\vk^\prime}
\left[1-\frac{v_{\vk^\prime}}{v_\vk}\frac{\tau(\epsilon_{\vk^\prime})}
{\tau(\epsilon_\vk)}\cos(\vartheta)\right]\frac{1-n_F(\epsilon_{\vk^\prime})}
{1-n_F(\epsilon_\vk)},
\]
where $\vartheta$ is the angle between $\vk$ and $\vk^\prime$.
The solution for $\tau(\epsilon_\vk)$ determines the electron
momentum relaxation time. 
When the scattering of quasiparticles from long-wavelength acoustic phonons is quasielastic, $\Omega_{q,n}\ll |\mu|$,
we can set $\tau(\epsilon_{\vk^\prime})=\tau(\epsilon_\vk)$
for the right-hand side of the above integral equation;
this is equivalent to the ``test particle approximation.''\cite{bannov}  
 The current density,
 ${\bf j}=-(e/\mathcal{A})\sum_\vk \mathbf{v}_\vk f(\vk)$,
points parallel to the electric field direction and has the magnitude
\[
 j=\frac{e^2 E}{\mathcal{A}}\sum_\vk v_\vk^2\cos^2(\alpha)\, 
\tau(\epsilon_\vk) [-\partial_\epsilon n_F(\epsilon_\vk)] .
\]
Using $v_\vk=v_k$ and performing the angular integration,
we arrive at the phonon contribution to the resistivity 
quoted in Eqs.~\eqref{eq:rho} and \eqref{momentum}.

\section{Transport Eliashberg function}
\label{appc}

We here give the analytical result for the full transport 
Eliashberg function ${\cal F}$ defined in Eq.~\eqref{eq:F}. 
Some straightforward yet tedious algebra
allows to perform the $\theta$-integration.
We find the (lengthy) result
\begin{widetext}
\begin{eqnarray*}
{\cal F}_{k,n,s}^{(\nu)}(q) &=&
\frac{2\left|A_{k,q,n,s}^{(\nu)}\right|}{\pi(A_2 W)^2}
\frac{\Theta\left(Q_{k,q,n,s}^{(\nu)}+k-q\right)
\Theta\left(q-\left|Q_{k,q,n,s}^{(\nu)}-k\right|\right)}
{\left[\left(q^2-(Q_{k,q,n,s}^{(\nu)}-k)^2\right)
\left((Q_{k,q,n,s}^{(\nu)}+k)^2-q^2\right)\right]^{1/2}}\,
\Theta\left(\left|A_{k,q,n,s}^{(\nu)}\right|-\Delta/2\right)\\ 
&\times&
\sum_{s'} \Theta\left(s'A_{k,q,n,s}^{(\nu)}\right)
\left[1-\left(1-\frac{\nu\Omega_{q,n}}{A_{k,q,n,s}^{(\nu)}}\right)
\frac{\left(Q_{k,q,n,s}^{(\nu)}\right)^2+k^2-q^2}{2k^2}\right] \left.
\left| M_{\vk,\vq,n}^{s',s}\right|^2\right|_{\theta_0},
\end{eqnarray*}
\end{widetext}
where we use the notations 
\begin{eqnarray*}
A^{(\nu)}_{k,q,n,s}&=& s\sqrt{(\Delta/2)^2+(A_2W k)^2} +
\nu\Omega_{q,n},\\
Q_{k,q,n,s}^{(\nu)} &=&
\frac{\sqrt{\left(A_{k,q,n,s}^{(\nu)}\right)^2-(\Delta/2)^2}}{A_2 W},
\end{eqnarray*}
and the polar angle $\theta=\theta_0\in [0,\pi]$ follows from 
\[
\sqrt{k^2+q^2+2kq\cos\theta_0} = Q_{k,q,n,s}^{(\nu)},
\]
fixing the polar angle between $\vk$ and $\vq$ in the matrix element $M$.


\begin{thebibliography}{99}
\bibitem{footcorr}
We mention that Eq.~(8) in Ref.~\onlinecite{giraud} contains a mistake: in the denominator of the expression for $C^{-1}$, one should replace $\kappa_l^2\kappa_t\to \kappa_l\kappa_t^2$. This implies $C\simeq 0.78$ instead of the quoted value $C\simeq 1.20$ in Ref.~\onlinecite{giraud}. Fortunately, this mistake has no consequences for the conclusions reached in Ref.~\onlinecite{giraud}.
\bibitem{foot}
In principle, the self-energy is off-diagonal in the $ss'$ indices, but
for the rate $\Gamma$, only the diagonal element with $s=s'$ is needed.
We also use the fact that $\Sigma$ then is $\upsilon$-independent.
\end{thebibliography}
\end{document}